\documentclass[times, 10pt,twocolumn]{article}
\usepackage{latex8}
\usepackage{times}
\usepackage{url}
\usepackage{epsfig}

\begin{document}

\title{High Performance Power Spectrum Analysis Using a FPGA Based Reconfigurable Computing
Platform}

\author{Yogindra Abhyankar, Sajish C, Yogesh Agarwal\\
Hardware Technology Development Group \\ Centre for Development of
Advanced Computing \\
Pune 411 007, India\\ yogindra@cdac.in\\
\and
C.R. Subrahmanya, Peeyush Prasad \\
Department of Astronomy and Astrophysics \\ Raman Research Institute \\ Bangalore 560 080, India\\
 crs@rri.res.in\\
}

\maketitle
\thispagestyle{empty}

\begin{abstract}
   Power-spectrum analysis is an important tool providing critical
   information about a signal.  The range of applications includes
   communication-systems to DNA-sequencing. If there is interference
   present on a transmitted signal, it could be due to a natural cause
   or superimposed forcefully.  In the latter case, its early detection
   and analysis becomes important.  In such situations having a small
   observation window, a quick look at power-spectrum can reveal a great deal
   of information, including frequency and source of interference.

In this paper, we present our design of a FPGA based
reconfigurable platform for high performance power-spectrum
analysis.  This allows for the real-time data-acquisition and
processing of samples of the incoming signal in a small time
frame.  The processing consists of computation of power, its
average and peak, over a set of input values.  This platform
sustains simultaneous data streams on each of the four input
channels.
\end{abstract}

\Section{Introduction} The concept and use of power spectrum of a
signal is fundamental in engineering - in communication
systems, microwave and radars. Recently, it is also being used in
diverse applications such as gene identification. In a typical
transmit-receive system, if the received signal is pure and as
expected, no filtering is required. However, on the other-hand,
any interference overriding the received signal may require
certain analysis in order to know more about the interference.  As
the interference tends to pump additional power in the received
waves, the power becomes a useful criterion for such an analysis.
Using the reverse-engineering techniques, the excess power
information with the incoming signal may help in finding the
characteristics of the interface such as frequency, source etc.

A power spectrum \cite{bingham:modern} is a representation of the
magnitude of the various frequency components of a signal.  By
looking at the spectrum, one can find how much energy or power is
contained in the frequency components of the signal.  Analysis or
evaluation of the power spectrum is one of the ways of isolating
noise.

There are a couple of techniques for generating the power
spectrum.  The most common one is by using the Fourier transform
\cite{brigham:fft}.  The other techniques such as the wavelet
transform or the maximum entropy method can also be used.

Experimentally, power spectrum can be determined in three ways:
(1) Using a spectrum or signal analyzer - a commercial instrument
\cite{national:instr} dedicated for displaying the real time power
spectra (2) Using a microcomputer based add-on signal analyzer
card, or (3) by digitizing experimental data and performing a Fast
Fourier Transform (FFT) on a desktop machine.

In terms of cost and complexity, the above-mentioned three options
are in the descending order, while considering the flexibility,
they are in the ascending order. Dedicated analyzers are some times
used, however they may not be cost effective, flexible or
competent enough, to extract the interference related information
when the observation window is short.

In general, the second option provides additional flexibility,
especially when the Field Programmable Gate Array (FPGA) is used.

In this paper we present our design of a very powerful
reconfigurable computing based design for solving complex signal
functions and real-time analysis.  Although this works as an
add-on card for a workstation, it is extremely powerful, flexible
and relatively cost effective.  The power spectrum analysis uses
modules developed by us for multi-channel data acquisition and
several signal processing operations performed simultaneously on
four data channels.

The FPGA based solution allows for the real-time acquisition and
processing of samples of the incoming signal. After the data
acquisition and analysis, the data is passed to the host, based on
the selected options.

Our card sustains simultaneous data streams on each of the four
channels for complex algorithms.

We begin this paper by briefly discussing the mechanics behind the
power spectrum analysis.  Section 3 outlines Reconfigurable
Computing and the card used for this work. In Section 4 and 5, we
discuss the scheme used for our implementation of power spectrum
analysis on the FPGA based reconfigurable hardware and the
experimental setup respectively. Finally, we summarize this paper
and indicate some directions for future improvements.
\Section{Power Spectrum Analysis} It is very difficult to detect
noise or interference if present in the input signal by merely
observing the time domain samples. However, by mapping the signals
\cite{rabiner:dsp} in frequency domain, the analysis and detection
of such signals becomes easy. The signal processing technique, in
particular the FFT plays an important role.  In 1965, it was
practically used by J.W. Cooley and J.W. Tukey of Bell Labs to
filter    the noisy signals.  This divide and conquer technique
for a set of N elements reduces the number of complex
multiplications to an order of N * $log_2$ N from $N^2$ otherwise
required by the Discrete Fourier Transform (DFT).

The power spectrum analysis uses FFT to represent the magnitude of
various frequency components of a signal.  By observing the
spectrum, one can find how much energy or power is contained in
the different frequency components of the signal.  Analysis of the
power spectrum allows isolating noise and provides information
related to its source.
\Section{Reconfigurable Computing (RC)} RC \cite{scott:recinfig}
explores the HW/SW solutions where the underlying hardware is 
flexible and is modified at runtime under software control to
accelerate an application. Predominantly, RC uses FPGA, a VLSI
chip whose hardware functionality is user-programmable. Putting
FPGAs on a PC add-on card or motherboard allows FPGAs to
serve as compute-intensive co-processors.   It is realized that
considerable acceleration may be achieved by targeting algorithms
in these application-specific, dynamically programmable flexible
parts.

Reconfigurable Computing- the paradigm to accelerate applications
using programmable hardware has sufficiently matured.  Now, HPC
community is looking towards this technology to further enhance
the power of clusters for supercomputing needs.

The following subsections summarizes the Reconfigurable hardware
and the system software used in this experimentation.
\SubSection{RC card} It is a FPGA based card
\cite{cdac:initiatives} that can be plugged to a host computer via
the 64-bit, 66 MHz PCI bus. This card has two Xilinx FPGAs
\cite{xilix:data}. Out of these, the larger device, XCV800 is
used as a compute engine implementing the application logic. The other FPGA is a XCV300 device
that holds the PCI controller and logic to control other devices.
When plugged into a PCI slot, the RC card can be assumed to work
as a co-processor to the host. Figure 1 shows the RC card block
diagram.
\begin{figure}
\centering \epsfig{file=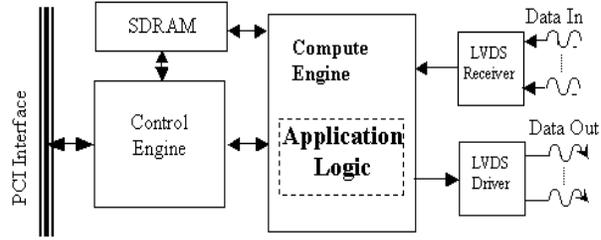,width=8 cm,height=3.5 cm}
\caption{RC card block diagram}
\end{figure}
There is an on-board 128MB of SDRAM and 1MB of ZBT RAM.  The SDRAM
is useful for storing input, intermediate and final results.  The
ZBT is suitable for applications where caching is required.  The
card supports DMA operations.

Input and output data to the card may be supplied from the host
using the PCI interface or it can directly come to the card using
the LVDS interface \cite{national:lvds}.   LVDS allows a high
speed data transfer in excess of 1 Gbps.

The system software interface for the RC card is implemented over
Red Hat Linux operating system.  It provides all the basic
functionalities in terms of the data transfer and card control
irrespective of the intended application.  The device driver
performs resource management and services to allocate/free DMA
buffers.  The system software also provides basic services to
configure, setup/free resources, send input data, receive output
data, initiate computation etc.
\Section{Power Spectrum Analyzer on RC}
\begin{figure*}
\centering \epsfig{file=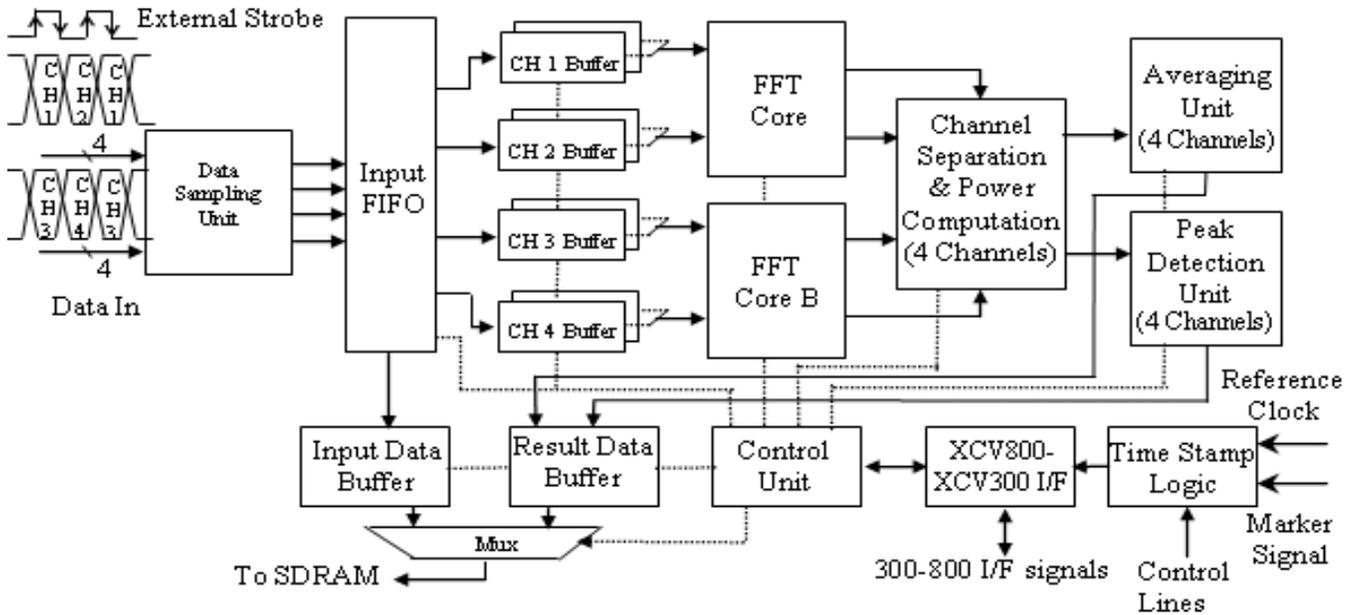,height=3.5 in, width= 7 in}
\caption{Power spectrum analyzer implemented on compute FPGA}
\end{figure*}
The power spectrum analyzer application has mainly two components:
the one running on the host system and the other running on the RC
card attached to the host.  The host controls the initial setup of
the application.  The raw input data is pre-processed by the RC
card, and power, average-power and peak-power values determined.
The host performs post-processing and other operations on the
processed data generated by the RC. This is required to complete
the power spectrum analysis.

As shown in figure 1, the input LVDS data-streams are handled by
the on-board receivers to provide compatible signals for the
compute engine.

The power spectrum computation block that resides in the XCV800
compute FPGA is shown in figure 2.  It consists of six main
components: Input Sampler and buffers, Multi-channel FFT units,
Channel Separator and Power computation unit, Average and Peak
Power Computational unit, Time-stamping and control, and the
XCV800-XCV300 interface.  In the following subsections, we
describe these components of the application.
\SubSection{Input Sampler and buffers} The spectrum analyzer
application requires four LVDS channels as inputs, each having a
4-bit data width.  However, there are only eight dedicated
differential lines for the channels.  The channels are
time-multiplexed in pairs, i.e. channel-1 and channel-2 goes on
four lines, while channel-3 and channel-4 on the remaining four
lines. A clock, serving as a strobe is provided. The data to the
sampler unit, changes on positive and negative edges of this
clock.

The channel-multiplexed input data is passed to the Sampler unit,
de-multiplexed and forwarded to channel buffers as well as to the
input-data-buffers.  The data from the channel buffers are input
to the FFT block, while the data from the input-data-buffers are
stored in the SDRAM.

The channel buffering is necessary to collect a block of data
before the FFT computation.  By using buffer pairs at each FFT
input, the data is read and processed by the FFT unit in parallel
to the input data streamed by the host in the other buffer.  When
the FFT core finishes processing the current input data, the
memory banks are swapped and the data load and computation
continues on the alternate memory bank.
\SubSection{Multi Channel FFT} This block uses two, 256-point
complex-FFT units from Xilinx CoreGen library, working in parallel
on the four input data channels. Instead of using them for complex
FFT computation having real and imaginary inputs, they are used
for processing two real data streams.  The units calculate complex
FFT according to the following equation:
\begin{equation}
X(k)  =  \frac{1}{256 \cdot s} \sum x(n)e^{\frac{-jnk2\pi}{256}}
\end{equation}
Where,

x(n) is the input sequence n = 0,1,2,....255;

X(k) is the output sequence k = 0,1,2 ...255;

s is the scaling factor adjusted to 1;
\SubSection{Channel Separation and power Computation} The channel
separation and power computation block separates the FFT values
for the two real channels from the complex FFT values, and
computes power for each channel.

As a result of the complex FFT, real and imaginary values are
obtained in the frequency domain. If the obtained values are
Re[256] and Im[256], the two channels are separated using the
following set of equations:
\begin{eqnarray*}
     CH1\_real[N] & = & (Re[N] + Re[256-N])/(2) \\
 CH2\_real[N] & = & (Im[N] - Im[256-N])/(2) \\
     CH1\_imag[N] & = & (Im[N] + Im[256-N])/(2) \\
     CH2\_imag[N] & = & (Re[256 -N] - Re[N])/(2)
\end{eqnarray*}
Similar equations hold good for channel 3 and 4.

The power values are calculated for each channel as per the
following equation:
\begin{eqnarray*}
         CHx\_pwr[N] = CHx\_real[N]^2  +  CHx\_imag[N]^2
\end{eqnarray*}

Where x represents channel number. The power values are positive, 32-bit
values, stored internally in Block RAMs.


\SubSection{Average and Peak Power Computation} The computation of
the average of the power values and the peak power values is done
in this block.  The computed peak and average values are stored
into the SDRAM.  Averaging of the results over a small period is
done, to enable the host software to read the results in parallel.
The Power Spectrum values are averaged over a period of 128 Blocks
(1Block = 256 points); along with averaging, the peak values
observed at each frequency point are stored.  All the results
obtained are written into the SDRAM. This is called as one Short
Term Accumulation (STA) cycle.

\SubSection{Time Stamp and Control} Time stamp and control block
has two 32-bit counters, Timestamp and Marker. These counters are
used for time stamping the input data, and operate on a reference
clock and a marker signal provided as input.  The Timestamp
counter runs on the reference clock and is reset on every marker
pulse. The Marker-counter increments with every marker pulse and
is reset with a reset given to the XCV800. These count values are
updated at the very instant a first data comes in a new cycle (1
Cycle = 128 STAs) and is given to the host.

\SubSection{XCV800-XCV300 Interface } The XCV800-XCV300 interface
allows communication between the compute and control engine.
There are a set of control and data lines, a set of registers and
a well-defined protocol that allows communication through the
interface.
\Section{Experimental Setup} The experimental setup is shown in
Figure 3, where the RC card having LVDS input capturing
capabilities is attached to a PCI based host.  Since the actual
input for the experimental setup is available in the form of
RS-422 signals, a small signal converter board for RS-422 to LVDS
is designed and connected at the RC card input.

\begin{figure}
\centering \epsfig{file=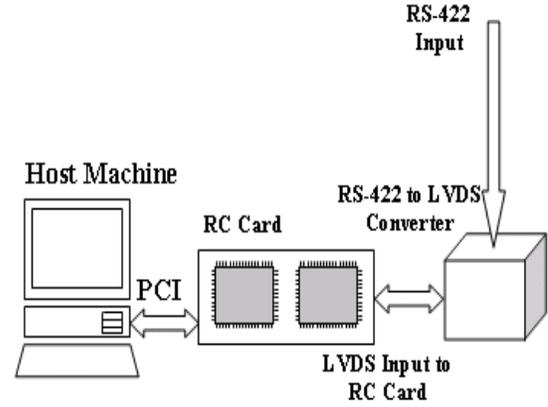,width=8.5 cm,height=6 cm}
\caption{Experimental setup}
\end{figure}

As mentioned in section 4, the code running on the host uses a set
of commands to control and initiate the application on the RC
card.  First of all, XCV800, the compute engine device is
configured.  After configuration, the device is given a reset.

The SDRAM on the card is used as two circular buffers, one for the
input data and the other for the result.  The size of each
circular buffer is set using the SETUP command.  The starting
address for input is kept fixed to the $1^{st}$ location of SDRAM;
similarly, the end address for results is kept fixed to the last
location of SDRAM.  Using the SETUP command, we set the $1^{st}$
location of the last block address for the input area and starting
block address for the result area.

Once the address is setup, the START command is issued to start
the compute engine. The Status, Timestamp and
Marker count registers are polled to control the application. The
Status register keeps a track of the SDRAM address, where the
application is currently writing the results.  The Timestamp
register indicates the current timestamp counter value.  The
Marker count register indicates the current marker counter value.

The processing of the data is stopped by issuing a STOP command.
With this, the application neither processes the data nor writes
to the SDRAM until a START command is issued.

Giving START after a STOP will restart the acquisition and
computation of data, and write results and input-data to the SDRAM.
These data values are written to the SDRAM from the starting addresses 
provided by the SETUP command.  The values
of timestamp and marker counts, before the STOP command and
after the START command indicates the time interval during which
the data was not processed.

Computed values of average power and peak power for all the four
channels are stored in the SDRAM. The averages are stored in the
first 256 locations followed by the peaks in the next 256
locations. The average power is stored as a 32-bit value.  The
32-bit peak power information carries the peak power value and the
corresponding block index where the peak has occurred.

The host software read the results from the SDRAM in parallel
while the application is running. The software synchronizes itself
to the application by polling the status register and performing a
DMA for reading out the results.  We found that when a large DMA
is done in parallel with the application, some part of the input
data is over-written. By experimenting with various DMA sizes, an
optimal DMA size of 4K was obtained that doesn't cause this data
loss.

After reading out the results from the RC card, the host performs
graphical data analysis with numerous power-frequency plots.

\Section{Results and Discussions} The hardware modules -
Input-sampler-buffer, channel-separator-power-computation,
Average-Peak power-computation, and Time-Stamping-Control are all
written in VHDL language, simulated using ModelSim 5.8 simulator
and synthesized using Xilinx ISE 5.1 tool.  All the designed
modules were optimized and runs at 66 MHz.  The 256-point Complex
FFT CoreGen component from Xilinx is instantiated and used along
with other modules. In this application, a single 256-point
complex FFT component was used to emulate two parallel real FFT
blocks.   For the 256-point FFT an average of 3 clock cycles are
required to calculate one FFT value.  Therefore our design can
sustain input data rates up to 22 MHz per channel.

We have also examined the reconfigurability of this card, by
selectively putting independent bit files for the average power or
peak power in the compute engine as per the user requirement.

The complete application for Average power implemented on a XCV800
compute FPGA utilizes around 80\% of slices and 92\% of block RAM.
The application with Peak power computation utilizes around 83\%
of slices and 92\% of block RAM.

One can easily port this application on a Xilinx Virtex-2Pro or
Virtex-4 device with a possibility of putting more than 8 FFT
cores and multiple power computation units, enhancing the
performance by many folds.  Here, we will have an added advantage
of having inbuilt LVDS signaling.

\Section{Conclusion} In this paper, we have presented a novel
application of reconfigurable computing for the detection of
interference using power spectrum analysis.  It uses in-house
developed modules along with the Xilinx FFT core.  The application
can also be reconfigured for computation of average power or peak
power based on the requirement.

Our design sustains simultaneous real-time data streams on each of
the four input channels.


\bibliographystyle{abbrv}
\bibliography{power_ref}

\end{document}